\documentclass[aps,prl,twocolumn,superscriptaddress,showpacs]{revtex4}

\usepackage{graphicx}
\begin{document}
\title{Clustering of Fermi particles with arbitrary spin}
\author{Andr\'as Csord\'as}
\email{csordas@tristan.elte.hu}
\affiliation{Research Group for Statistical Physics of the 
             Hungarian Academy of Sciences,
             P\'azm\'any P. S\'et\'any 1/A, H-1117 Budapest, Hungary}
\author{P\'eter Sz\'epfalusy}
\email{psz@galahad.elte.hu}
\affiliation{Department of Physics of Complex Systems, E\"otv\"os University,
             P\'azm\'any P. S\'et\'any 1/A, H-1117 Budapest, Hungary}
\affiliation{Research Institute for Solid State Physics and Optics,
             P. O. Box 49, H-1525 Budapest, Hungary}
\author{\'Eva Sz\H oke}
\email{szoke@ludens.elte.hu}
\affiliation{Budapest University Catholic Secondary School,
             Szab\'o Ilonka u. 2-4, H-1015 Budapest, Hungary}
\date{\today}
\begin{abstract}
A single $l$-shell model is investigated for a system of fermions
of spin $s$ and an attractive $s$-wave, spin channel independent,
interaction. The spectra and eigenvectors are determined exactly for
different $l$, $s$ values and particle numbers $N$. As a generalization
of Cooper pairing it is shown that when $N=\mu(2s+1)$, $\mu=1,2,\ldots,2l+1$,
the ground state consists of clusters of $(2s+1)$ particles.
The relevance of the results for more general situations including the
homogeneous system is briefly discussed. 
\end{abstract}
\maketitle

The experimental realizations of highly degenerate atomic 
Fermi gases confined in traps \cite{cataliotti98,ohara02,granade02,%
schreck01,demarco01,demarco99,truscott01}
have  increased the activity in the
theoretical investigation of such systems, which had already started
earlier in view of the hope of future experiments. 
In particular many aspects
of the possible transition to the superfluid phase have been studied
in the case of the attractive interaction between the particles
\cite{heiselberg00,heiselberg02,heiselberg01,grasso03,bulgac02,bruun02,%
bruun02a,chiofalo02,viverit01,holland01,houbiers97,stoof96,houbiers99,%
bruun99,kagan96,petrov03,baranov98a,leggett80,modawi97,mackie00}.
For instance, the critical temperature for the Cooper pairing of atoms
with spin values higher than $1/2$ has been investigated 
\cite{modawi97,heiselberg00} based on the Gorkov-Melik-Barkhudarov approach
to calculate corrections to the BCS theory due to 
induced interaction \cite{gorkov61}. The result is that the
critical temperature of the BCS theory is multiplied by $(4e)^{\nu/3-1}$,
where $\nu$ denotes the number of fermion species ($\nu=2s+1$) 
\cite{heiselberg00}.
Accordingly, if $s = 1/2$ (the case treated by Gorkov-Melik-Barkhudarov) 
the critical temperature
is decreasing, while it is increasing for $s \ge 3/2$.
In the special case $s=1$ no such correction exists \cite{modawi97}.

The aim of the present paper is to propose and demonstrate that the 
superfluid ground state can be different from that given by the
BCS pairing theory if the spin of the atoms $s>1/2$. Namely, we
suggest that instead of Cooper pairs the particles can create 
clusters of zero total spin value containing $2s+1$ particles.
This can occur in optical traps where the spins of the particles
can freely rotate (Alternatively, one can think about $2s+1$ hyperfine
states. In the following we use always the terminology of spin). 
We do not attempt to make any comparison with 
experimental results in this paper and choose the possible simplest
model to make the presentation free from approximations. We hope that
in future experiments the clustering effect can be revealed.
We treat the case when the number of species $(2s+1)$ is even.
It is straightforward to extend our considerations for odd number of species.
Note that in the spin-1 Fermi superfluid the pair correlation has been
investigated in detail by Modawi and Leggett \cite{modawi97}. 

We assume that the trap potential is spherically symmetric.
The investigation will be done at zero temperature and is restricted
to the subspace of a given $l$-shell filled with particles of number
$N$ ($0<N<(2l+1)(2s+1)$). The interaction between the particles in the
open shell is described by a spin independent contact potential representing
$s$-wave scattering with a negative scattering length:
\begin{equation}
H_{int}=-{\lambda \over 2}\sum_{i,j=1 \atop i \ne j}^N
    \delta(\mathbf{r}_i-\mathbf{r}_j).
\label{eq:firstqop}
\end{equation}
The one particle normalized wave functions in the open shell are given by
\begin{equation}
\Psi_{n,l,m_l,s,m_s}(r,\vartheta,\varphi,\sigma)=
R_{n,l}(r)Y^l_{m_l}(\vartheta,\varphi)\chi^s_{m_s}(\sigma),
\label{eq:onepf}
\end{equation}
Here $\sigma$ is a discrete spin variable, which can take
$(2s+1)$ different values. Spin eigenfunctions are ortho-normalized 
according to
\begin{equation}
\sum_\sigma \chi^s_{m_s}(\sigma) \chi^{s'}_{m'_s}(\sigma)=
\delta_{s,s'}\delta_{m_s,m'_s}
\end{equation}
The functions (\ref{eq:onepf}) are eigenfunctions of the one-particle
Hamiltonian, which contains besides the trap potential also the average field 
of the closed shells. The problem of diagonalizing the operator 
(\ref{eq:firstqop}) can be solved exactly on the fixed basis (\ref{eq:onepf}). 

Let us denote by $a_{m_l,m_s}$ the operator which annihilates a particle
with quantum numbers ($n$,$l$,$m_l$,$s$,$m_s$). 
In second quantization 
(\ref{eq:firstqop}) can be written as
\begin{eqnarray}
\hat{H}_{int} \equiv  \hat{h}E_0/\pi =&&
-{E_0 \over 2}\sum_{m_1,m_2,m_3,m_4}\sum_{\nu_1,\nu_2}
f_{m_1,m_2,m_3,m_4} \nonumber\\
&& \times a^+_{m_1,\nu_1}a^+_{m_2,\nu_2}a_{m_4,\nu_2}a_{m_3,\nu_1},
\label{eq:genham}
\end{eqnarray}
where $E_0$ is the characteristic energy
\begin{equation}
E_0=\lambda \int_0^\infty |R_{n,l}(r)|^4 r^2\, dr,
\end{equation}
and the coefficient $f$ is given in terms of 
the Wigner-$3j$ symbols.

The Hamiltonian (\ref{eq:genham}) can be converted into
\begin{equation}
\hat{h}={[l] \over 8}\hat{N}-{[l]^2 \over 8}\sum_{L=0 \atop L:even}^{2l}
\left(\begin{array}{ccc}
l & l & L \\
0 & 0 & 0
\end{array}\right)^2 \hat{B}_L^2,
\end{equation}
where the abbreviation $[a]\equiv (2a+1)$ is used and the scalar ``square''
\begin{equation}
\hat{B}_L^2=\sum_{M=-L}^L (-1)^{L-M}\hat{B}_{L,M}\hat{B}_{L,-M}
\end{equation}
of the irreducible tensor operators 
\begin{eqnarray}
\hat B_{L,M}=\sum_{m=-l}^l\sum_{\nu=-s}^s &&(-1)^{l-m}\sqrt{[L]}
\left(\begin{array}{ccc}
l & l & L \\
m & M-m & -M
\end{array}\right)\nonumber\\
&&\times a^+_{m,\nu}a_{m-M,\nu}
\end{eqnarray}
has been introduced. $\hat{N}$ stands for the particle number operator.
We note that in cases of $l=1$ and $l=2$ the Hamiltonian 
can be expressed in terms of the operators $\hat{N}$, $C_2(SU(3))$,
$C_2(SO(3))$ and $\hat{N}$, $C_2(SU(5))$, $C_2(SO(5))$, respectively,
where $C_2(G)$ denotes the quadratic Casimir operator of the group $G$ 
\cite{unp_our}.

We have calculated the full exact spectra and the
eigenvectors of $\hat{h}$ for $s=1/2$, $l=1,2,3$ and for $s=3/2$,
$l=1,2$ using some Mathematica and Fortran programs. Unfortunatelly,
the size of the Hilbert-space (in the $L_z=0$ $S_z=0$ or $S_z=1/2$ subspace
too) grows drastically. For higher $s$ we also have performed 
calculations  for $l=1$, $s=5/2,7/2$ in the $L=0$ subspace for even $N$.
The Young tableau technique has also been used to analyse the model.
The details will be published elsewhere \cite{unp_our}.
Results for the ground state energies up to the half filled
shell can be found in Tables \ref{table:l_1}. and \ref{table:l_2}. 
Above the half filling
energies and eigenvectors can be obtained by particle$\leftrightarrow$hole
transformation. If one exact eigenstate has $N$ particle in the $l,s$-shell
with energy $E(N)$ then there is an other exact eigenstate 
for particle number $N'=N_t-N$ ($N_t\equiv(2s+1)(2l+1)$) 
with energy
\[
E(N_t-N)=E(N)+(2l+1)s(N_t/4-N/2), \quad N\le N_t/2
\]  

It is enlightening to discuss the structure 
of the ground state for even number of
particles in coordinate representation as well. Let us start with
the two particle states. The wave function is as follows:
\begin{equation}
\psi(1,2)=\varphi(\mathbf{r}_1,\mathbf{r}_2)\, {}^{2S+1}\chi(\sigma_1,\sigma_2)
\label{eq_tpwf}
\end{equation}
The function $\varphi$ is symmetric and is the eigenfunction of the
angular momentum operator with $L=0$, while the spin functions
are antisymmetric. The latter is ensured if the total spin $S$ takes the values
$S=0,2,\ldots,2s-1$, which makes the ground state $2s$ times degenerate.
The degeneracy is lifted and a spin multiplet arises, if the $s$-wave
scattering lengths were different in different spin channels.
In case of $s=1/2$ it is found that the exact ground states are ``pair states''
for $N=2,4,\ldots,2(2l+1)$, i.e., their wave functions
 are the antisymmetrized products
of singlet pair wave functions
\begin{equation}
\Psi(1,2,\ldots,N)=\hat{A}\prod_{i=1}^{N/2}\psi(2i-1,2i).
\label{eq_pairwf}
\end{equation}
The operator $\hat{A}$ stands for antisymmetrization.
The situation is different when $s>1/2$. Namely, instead of pair states
we find that $2s+1$ particle clusters are created when
$N=\mu(2s+1)$, $\mu=1,2,\ldots,2l+1$.  
The wave function reads as
\begin{eqnarray}
\Psi(1,2,\ldots,N)=\hat{A}{\prod_i}'&&
\varphi(\mathbf{r}_i,\mathbf{r}_{i+1},\ldots,\mathbf{r}_{i+2s})
\nonumber\\
&&\,{}^1\chi(\sigma_i,\sigma_{i+1},\ldots,\sigma_{i+2s})
\label{eq_gswf}
\end{eqnarray}
The prime on the product sign in Eq.~(\ref{eq_gswf}) 
means that $i$ extends to the values as follows:
\begin{equation}
i=(2s+1)k+1, \qquad k=0,1,\ldots,\mu-1
\end{equation}
${}^1\chi$ denotes the singlet spin function representing the Slater
determinant of the (2s+1) one particle spin funtions $\chi_{m_s}$
\begin{eqnarray}
&&{}^1\chi(\sigma_1,\sigma_2,\ldots,\sigma_{2s+1})=
{1 \over \sqrt{(2s+1)!}} \nonumber\\
&&\times\left|\begin{array}{cccc}
\chi_{-s}(\sigma_1)&\chi_{-s}(\sigma_2)&\ldots&\chi_{-s}(\sigma_{2s+1}) \\
\chi_{-s+1}(\sigma_1)&\chi_{-s+1}(\sigma_2)&\ldots&
           \chi_{-s+1}(\sigma_{2s+1}) \\
\vdots & \vdots & \ddots& \vdots\\
\chi_s(\sigma_1)&\chi_s(\sigma_2)&\ldots&\chi_s(\sigma_{2s+1})
\label{eq:singlsf}
\end{array}\right|
\end{eqnarray}
In Eq.~(\ref{eq_gswf})
$\varphi$ is a completely symmetric function for the exchange of any pair
of particles. Furthermore, it is an eigenfunction of the
orbital angular momentum operator of the $2s+1$ particles with eigenvalue 
zero.

Eq.~(\ref{eq_gswf}) means that the states can be 
created by repeated application
of a creation operator to the vacuum 
$\left(\hat{Q}_{2s+1}^{(l)}\right)^\mu|0\rangle$.
We have derived this operator, but we do not present here in general form.
For $s=1/2$ the operator $\hat{Q}^{(l)}_2$ is the pair creation operator 
\begin{equation} 
Q_2^{(l)}={1 \over 2}\sum_{m=-l}^l
\sum_{\nu=-1/2}^{1/2}(-1)^{(l-m+1/2-\nu)}a^+_{m,\nu}a^+_{-m,-\nu}
\label{eq_q2}
\end{equation}
and for the ground state energy the simple relationship is obtained
for $s=1/2$:
\begin{equation}
E_0={2l+1\over 8\pi}N, \qquad N(\mbox{even})\leq 2(2l+1)
\label{eq_groundhalfspin}
\end{equation}
Furthermore, for $N=2$ Eq.~(\ref{eq_groundhalfspin}) gives also the
ground state energy for any $s$ values.

One can show that in the single $l$-shell model the cluster wave
function $\varphi$ in Eq.~(\ref{eq_gswf}) has the structure
\begin{equation}
\varphi(\mathbf{r}_1,\ldots,\mathbf{r}_{2s+1})=\hat{S}
\prod_{j=1}^{s+1/2}\varphi(\mathbf{r}_{2j-1},\mathbf{r}_{2j}),
\label{eq_cwf}
\end{equation}
where $\hat{S}$ denotes the operator for symmetrization and 
$\varphi(\mathbf{r}_1,\mathbf{r}_2)$ is the function defined in 
Eq.~(\ref{eq_tpwf}). It can be expressed in terms of the one particle
wave functions:
\begin{eqnarray}
\varphi(\mathbf{r}_1,\mathbf{r}_2)&=&R_{n,1}(r_1)R_{n,2}(r_2) \nonumber\\
&&\times \sum_{m=-l}^l (-1)^{l-m}Y_m^l(\Omega_1)Y_{-m}^l(\Omega_2).
\end{eqnarray}

\begin{figure}
\includegraphics[height=\linewidth,angle=270]{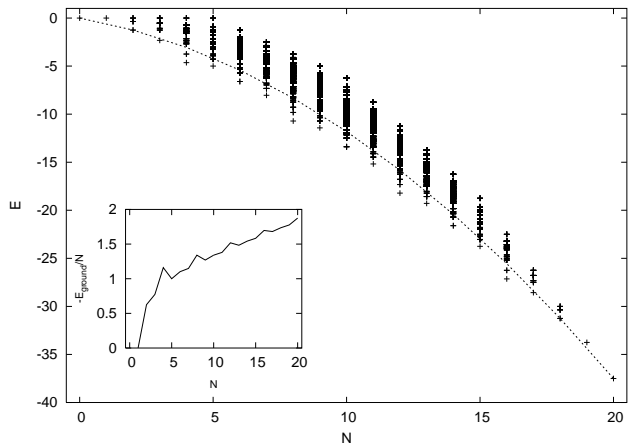}
\caption{\label{fig:1} The energy levels of the dimensionless
Hamilton operator $\hat{h}$ in case of $l=2$,
$s=3/2$. The dotted line (drawn everywhere to lead the eye) at even $N$
gives the average energy calculated using the pair wave function. 
The insertion shows the binding energy per particle as a function of the
number of particles.} 
\end{figure}

In Fig.~\ref{fig:1} besides the exact levels for $l=2$, $s=3/2$ 
the expectation value
of the Hamiltonian $\hat{h}$, Eq. (\ref{eq:genham}), with wave function 
(\ref{eq_pairwf}) is also shown.
As can be seen it provides the exact ground state energy only
for $N=2,(2s+1)(2l+1)-2,(2s+1)(2l+1)$. In the other cases the energy
obtained with the help of (\ref{eq_pairwf}) lies considerably higher
than the ground state energy. This is especially true when $N=\mu(2s+1)$,
and the wave function (\ref{eq_gswf}) applies. The binding energy
per particle has maximum at these particle numbers. 

For understanding the origin of the ground state wave function (\ref{eq_gswf})
one has to realize that
the relevant group of the model is $SU(2l+1)\otimes SU(2s+1)$.
With the usual decomposition one gets states characterized by quantum numbers
specifying the irreducible representations of 
$SU(2l+1)\supset SO(2l+1)\supset SO(3)$ and similarly for $SU(2s+1)$,
where the relevant quantum number is the total spin related to the
irreducible representations of $SU(2)\subset SU(2s+1)$. The seniority
is associated with $SO(2l+1)$, while the orbital angular momentum $L$
with $SO(3)$. The validity of the cluster wave function $\Psi$  can be best 
understood by studying first the structure of its spin function. For
$\mu=1$ the spin function is obviously invariant under the group
$SU(2s+1)$. It means that the generators of $SU(2s+1)$ commute with the
operator $\hat{Q}_{2s+1}^{(l)}$ introduced in the text above (\ref{eq_q2}). 
Since $\Psi$ is obtained by repeated
application of $\hat{Q}_{2s+1}^{(l)}$ it follows that $\Psi$ is also invariant
under $SU(2s+1)$ (consequently $S=0$). The corresponding Young tableau
consists of $\mu$ columns of length $(2s+1)$, which is known to specify
a one dimensional irreducible representation of $SU(2s+1)$ 
(see e.g. \cite{hamermesh64}).
The Pauli principle to be fulfilled by $\Psi$ requires that the corresponding
Young tableau of $SU(2l+1)$ must consist of $\mu$ rows, each of length
$(2s+1)$. It contains for each $\mu$ a one dimensional representation
of $SO(2l+1)$, which leads to $L=0$. The state is unambigously 
determined by choosing this representation of $SO(2l+1)$. We emphasize
that the symmetry of $\Psi$ is higher than just the rotational 
invariance in the coordinate
space and in the spin space. As a matter of fact the symmetry is the 
highest possible for particle numbers $N=\mu(2s+1)$.

In summary, our results are in harmony with the expectation
that the ground state of a system of Fermi particles with attractive
interaction is the possible most symmetric one with respect to the
exchange of the space coordinates of the particles. This feature 
manifests itself spectacularly, when the number of particles is the
multiple of the number of species $2s+1$. Though in this paper the
single $l$-shell model has been treated one can presume the validity
of the conclusions more generally. 

As a generalization one can take into account 
more than one shell in constructing
the cluster wave function $\varphi$ (configuration interaction). Note
that the totally antisymmetric spin functions 
${}^1\chi$, Eq. (\ref{eq:singlsf}), would not change. Another strategy
would be to treat first the homogeneous system and apply a local density
approximation for the trapped gas. The problem of the ground state of the
homogeneous system is, of course, interesting in itself. In this case $\varphi$
is translationally invariant. When $s=1/2$ the wave function $\Psi$,
Eq. (\ref{eq_gswf}), which coincides now with Eq.~(\ref{eq_pairwf}) is the well
known expression of the BCS ground state projected onto the $N$ particle
state and written in coordinate representation (see e.g. \cite{schrieffer64}). 
For the spin-1 Fermi gas, introduced by Modawi and Leggett \cite{modawi97},
we predict a ground state containing 
three particles,  clusters of zero spin and with a symmetric 
$\varphi(\mathbf{r}_1,\mathbf{r}_2,\mathbf{r}_3)$ function 
in the expression (\ref{eq_gswf}). 

As an example let us assume that four particles with spin $3/2$ 
are put outside the 
Fermi sea of noninteracting particles
as a generalization of the original Cooper problem 
for two particles (see for its discussion 
\cite{schrieffer64}).
One has to emphasize that even that part of the wave function,
which depends on the space coordinates
does not have the ``pair structure'' (\ref{eq_cwf}) in general, it is valid
only within the single $l$-shell model. One can use (\ref{eq_cwf}) 
as an ansatz, however,
in a homogeneous system and compare it with (\ref{eq_pairwf}) for the
ground state. Now $\varphi(\mathbf{r}_1,\mathbf{r}_2)$ is specified
by the requirement that it should satisfy the equation for a bound Cooper pair
(because of the $\delta$ function nature
of the two particle interaction one needs the usual regularization 
(see e.g. \cite{houbiers97})).
The result of a simple, but somewhat lengthy
calculation leads to:
\begin{equation}
E_{pairs}-E_{4cluster}={5\lambda\over V}
\left[1+{\varphi(\mathbf{0})\int d^3r\, \varphi(\mathbf{r})C(\mathbf{r})
\over C^2(\mathbf{0})} 
\right]+O({1\over V^2})
\label{eq_diff}
\end{equation}
where $E_{pairs}$ and $E_{4cluster}$ are the expectation values of the
Hamiltonian with the wave functions of form (\ref{eq_pairwf}) 
and (\ref{eq_gswf},\ref{eq_cwf})
respectively, using the interaction (\ref{eq:firstqop}). 
$V$ denotes the volume of the system. It has been used that 
$\varphi(\mathbf{r}_1,\mathbf{r}_2)=\varphi(|\mathbf{r}_1-\mathbf{r}_2|)$ 
can be taken real.  The
function $C(\mathbf{r})$ is defined as
$
C(\mathbf{r}_1-\mathbf{r}_2)=\int d^3r_3\, \, 
\varphi(|\mathbf{r}_1-\mathbf{r}_3|)
\varphi(|\mathbf{r}_3-\mathbf{r}_2|)
$.
As expected the difference (\ref{eq_diff}) 
is of $O(1/V)$. Since $\varphi$ and $C$ are
dominantly positive the result shows that the wave function (\ref{eq_gswf})
is energetically favored due to the fact that (\ref{eq_gswf}) is 
``more symmetric'' in space coordinates than (\ref{eq_pairwf}).
To get an energy gain of $O(1)$ a more accurate ground state wave
function is needed, which can account for a possible four particle bound state.
To find it goes beyond the scope of the present paper and is a part
of our planned investigations.

\begin{acknowledgments}
The present work has been partially supported by the Hungarian Research
National Foundation under Grant Nos. OTKA T029552 and T038202.
\end{acknowledgments}

\squeezetable
\begingroup
\begin{table}
\begin{tabular}{lccccccc}
\hline
N       &  2   &   4     & 6   & 8 & 10 & 12 & 14 \\
\hline\hline
$s=1/2$ & $-3/4$ \\
$s=3/2$ & $-3/4$  & $-33/10$& $-93/20$\\
$s=5/2$ & $-3/4$  & $-33/10$& $-153/20$ & $-48/5$\\
$s=7/2$ & $-3/4$  & $-33/10$& $-153/20$ & $-69/5$& $-327/20$
        & $-207/10$\\
$s=9/2$ & $-3/4$  & $-33/10$& $-153/20$ & $-69/5$& $-87/4$ & $-249/10$ &
        $-597/20$ \\
\hline
\end{tabular}
\caption{Ground state energies of the dimensionless Hamilton 
operator $\hat{h}$ for $l=1$ and for $N=0,2,\ldots,3(2s+1)/2$. 
\label{table:l_1}}
\end{table}
\endgroup

\squeezetable
\begingroup
\begin{table}
\begin{tabular}{lccccc}
\hline
N      &      2   &     4 &    6 &  8  &  10  \\
\hline\hline
$s=1/2$ & $-5/4$  & $-5/2$\\
$s=3/2$ & $-5/4$  & $-65/14$& $-185/28$ & $-75/7$ & $-375/28$\\
\hline
\end{tabular}
\caption{Ground state energies of the dimensionless Hamilton 
operator $\hat{h}$ for $l=2$ and for $N=0,2,\ldots,5(2s+1)/2$. 
\label{table:l_2}}
\end{table}
\endgroup

\end{document}